\documentclass[10pt,preprint]{IEEEtran}

\IEEEoverridecommandlockouts
\usepackage{xcolor}
\usepackage{graphicx}
\usepackage{booktabs}
\usepackage{array}
\usepackage{cite}
\usepackage{url}
\usepackage{tikz}
\usetikzlibrary{positioning}
\usepackage{scratch3}
\usepackage{algorithm}
\usepackage{algpseudocode}
\usepackage[hidelinks]{hyperref}

\newcommand{\Description}[1]{}

\newcommand{\tool}{\mbox{ScratchLens}}
\newcommand{\scratchlang}{Scratch}
\newcommand{\lfinal}{\ensuremath{L_{\mathit{final}}}}
\newcommand{\lframe}{\ensuremath{L_{\mathit{frame}}}}
\newcommand{\levent}{\ensuremath{L_{\mathit{event}}}}
\newcommand{\lmonitor}{\ensuremath{L_{\mathit{monitor}}}}
\newcommand{\ldefault}{\ensuremath{L_{\mathit{default}}}}
\newcommand{\SpectraSemanticCases}{26}

\newcommand{\SpectraRootCauseAccuracy}{1.000}

{26}

\newcommand{\SpectraRootCauseAccuracy}{1.000}

}

\newcommand{\MatrixUsablePairs}{444}
\newcommand{\MatrixTotalPairs}{544}
\newcommand{\MatrixQuarantined}{12}
\newcommand{\MatrixAmbiguous}{88}
\newcommand{\MatrixSpectraAccuracy}{1.000}
\newcommand{\MatrixSpectraOverall}{1.000}
\newcommand{\MatrixSpectraCoverage}{1.000}
\newcommand{\MatrixSpectraFalseEq}{0}
\newcommand{\MatrixSpectraPrecision}{1.000}
\newcommand{\MatrixSpectraRecall}{1.000}

\newcommand{\MatrixSpectraDiag}{0.987}
\newcommand{\MatrixSpectraFinalAccuracy}{0.721}
\newcommand{\MatrixSpectraFinalOverall}{0.721}

\newcommand{\MatrixSpectraFinalFalseEq}{124}

\newcommand{\MatrixSpectraNoporAccuracy}{0.955}
\newcommand{\MatrixSpectraNoporOverall}{0.955}

\newcommand{\MatrixSpectraNoporFalseEq}{0}

\newcommand{\MatrixSpectraNoporDiag}{0.987}
\newcommand{\MatrixBHashAccuracy}{0.579}
\newcommand{\MatrixBHashOverall}{0.579}

\newcommand{\MatrixBHashFalseEq}{17}

\newcommand{\MatrixBOpcodeAccuracy}{0.597}
\newcommand{\MatrixBOpcodeOverall}{0.597}

\newcommand{\MatrixBOpcodeFalseEq}{19}

\newcommand{\MatrixBTraceAccuracy}{0.608}
\newcommand{\MatrixBTraceOverall}{0.608}

\newcommand{\MatrixBTraceFalseEq}{43}

\newcommand{\MatrixBStructAccuracy}{0.590}
\newcommand{\MatrixBStructOverall}{0.590}

\newcommand{\MatrixBStructFalseEq}{0}

\newcommand{\MatrixBDynfiveAccuracy}{0.899}
\newcommand{\MatrixBDynfiveOverall}{0.899}

\newcommand{\MatrixBDynfiveFalseEq}{28}

\newcommand{\MatrixLlmGlmAccuracy}{0.969}
\newcommand{\MatrixLlmGlmOverall}{0.928}
\newcommand{\MatrixLlmGlmCoverage}{0.957}
\newcommand{\MatrixLlmGlmFalseEq}{9}

\newcommand{\MatrixLlmGlmDiag}{0.816}
\newcommand{\MatrixLlmQwenAccuracy}{0.959}
\newcommand{\MatrixLlmQwenOverall}{0.849}
\newcommand{\MatrixLlmQwenCoverage}{0.885}
\newcommand{\MatrixLlmQwenFalseEq}{15}

\newcommand{\MatrixLlmQwenDiag}{0.532}
\newcommand{\MatrixLlmKimiAccuracy}{0.982}
\newcommand{\MatrixLlmKimiOverall}{0.977}
\newcommand{\MatrixLlmKimiCoverage}{0.995}
\newcommand{\MatrixLlmKimiFalseEq}{8}

\newcommand{\MatrixLlmKimiDiag}{0.899}

\newcommand{\MatrixSpectraHardOverall}{1.000}
\newcommand{\MatrixSpectraHardFalseEq}{0}

\newcommand{\MatrixBStructHardOverall}{0.313}

\newcommand{\MatrixBDynfiveHardOverall}{0.871}
\newcommand{\MatrixBDynfiveHardFalseEq}{14}
\newcommand{\MatrixLlmGlmHardOverall}{0.951}
\newcommand{\MatrixLlmGlmHardFalseEq}{5}
\newcommand{\MatrixLlmQwenHardOverall}{0.755}
\newcommand{\MatrixLlmQwenHardFalseEq}{7}
\newcommand{\MatrixLlmKimiHardOverall}{0.982}
\newcommand{\MatrixLlmKimiHardFalseEq}{1}
\newcommand{\MatrixSpectraSingleOverall}{1.000}

\newcommand{\MatrixBStructSingleOverall}{0.751}

\newcommand{\MatrixLlmGlmSingleOverall}{0.915}
\newcommand{\MatrixLlmQwenSingleOverall}{0.904}
\newcommand{\MatrixLlmKimiSingleOverall}{0.975}
\newcommand{\MatrixRqfourBucket}{88}
\newcommand{\MatrixRqfourStaticDiff}{88}
\newcommand{\RqFourExposed}{14}
\newcommand{\RqFourRate}{0.159}
{--}
\newcommand{\MatrixTotalPairs}{--}
\newcommand{\MatrixQuarantined}{--}
\newcommand{\MatrixAmbiguous}{--}
\newcommand{\MatrixSpectraAccuracy}{--}
\newcommand{\MatrixSpectraCoverage}{--}
\newcommand{\MatrixSpectraFalseEq}{--}
\newcommand{\MatrixSpectraPrecision}{--}
\newcommand{\MatrixSpectraRecall}{--}

\newcommand{\MatrixSpectraFinalAccuracy}{--}
\newcommand{\MatrixSpectraNoporAccuracy}{--}
\newcommand{\MatrixSpectraNoporFalseEq}{--}
\newcommand{\MatrixBHashAccuracy}{--}
\newcommand{\MatrixBHashFalseEq}{--}
\newcommand{\MatrixBOpcodeAccuracy}{--}
\newcommand{\MatrixBOpcodeFalseEq}{--}
\newcommand{\MatrixBTraceAccuracy}{--}
\newcommand{\MatrixBTraceFalseEq}{--}
\newcommand{\MatrixBStructAccuracy}{--}
\newcommand{\MatrixBStructFalseEq}{--}
\newcommand{\MatrixBDynfiveAccuracy}{--}
\newcommand{\MatrixBDynfiveFalseEq}{--}
\newcommand{\MatrixRqfourBucket}{--}
\newcommand{\MatrixRqfourStaticDiff}{--}
}

\providecommand{\MatrixLlmGlmAccuracy}{--}
\providecommand{\MatrixLlmGlmFalseEq}{--}
\providecommand{\MatrixLlmGlmCoverage}{--}
\providecommand{\MatrixLlmKimiAccuracy}{--}
\providecommand{\MatrixLlmKimiFalseEq}{--}
\providecommand{\MatrixLlmKimiCoverage}{--}
\providecommand{\RqFourExposed}{--}
\providecommand{\RqFourRate}{--}
\providecommand{\MatrixSpectraOverall}{--}
\providecommand{\MatrixSpectraDiag}{--}
\providecommand{\MatrixSpectraNoporOverall}{--}
\providecommand{\MatrixSpectraNoporDiag}{--}
\providecommand{\MatrixSpectraFinalOverall}{--}
\providecommand{\MatrixSpectraFinalFalseEq}{--}

\providecommand{\MatrixBHashOverall}{--}

\providecommand{\MatrixBOpcodeOverall}{--}

\providecommand{\MatrixBTraceOverall}{--}

\providecommand{\MatrixBStructOverall}{--}

\providecommand{\MatrixBDynfiveOverall}{--}

\providecommand{\MatrixLlmQwenAccuracy}{--}
\providecommand{\MatrixLlmQwenFalseEq}{--}
\providecommand{\MatrixLlmQwenCoverage}{--}
\providecommand{\MatrixLlmGlmOverall}{--}
\providecommand{\MatrixLlmGlmDiag}{--}
\providecommand{\MatrixLlmQwenOverall}{--}
\providecommand{\MatrixLlmQwenDiag}{--}
\providecommand{\MatrixLlmKimiOverall}{--}
\providecommand{\MatrixLlmKimiDiag}{--}

\providecommand{\MatrixSpectraHardOverall}{--}
\providecommand{\MatrixSpectraHardFalseEq}{--}
\providecommand{\MatrixSpectraSingleOverall}{--}
\providecommand{\MatrixBDynfiveHardOverall}{--}
\providecommand{\MatrixBDynfiveHardFalseEq}{--}
\providecommand{\MatrixBStructHardOverall}{--}
\providecommand{\MatrixBStructSingleOverall}{--}
\providecommand{\MatrixLlmGlmHardOverall}{--}
\providecommand{\MatrixLlmGlmHardFalseEq}{--}
\providecommand{\MatrixLlmGlmSingleOverall}{--}
\providecommand{\MatrixLlmQwenHardOverall}{--}
\providecommand{\MatrixLlmQwenHardFalseEq}{--}
\providecommand{\MatrixLlmQwenSingleOverall}{--}
\providecommand{\MatrixLlmKimiHardOverall}{--}
\providecommand{\MatrixLlmKimiHardFalseEq}{--}
\providecommand{\MatrixLlmKimiSingleOverall}{--}

\begin{document}

\title{\tool: Lens-Parametric Behavioral Equivalence for Scratch Programs}

\author{
\IEEEauthorblockN{Yuan Si and Jialu Zhang$^{*}$\thanks{Corresponding author: Jialu Zhang.}}\\
\IEEEauthorblockA{University of Waterloo, Waterloo, Canada\\
\texttt{yuan.si@uwaterloo.ca}, \texttt{jialu.zhang@uwaterloo.ca}}
}

\maketitle

\begin{abstract}
Two \scratchlang{} programs can be syntactically far apart through renamed
variables, split scripts, extracted custom blocks, and reordered
initialization, while preserving the same behavior. A one-block edit, such as
replacing a blocking broadcast with an asynchronous one, can create
divergences that surface only under specific schedules. Behavioral
equivalence for such programs is central to automated feedback, grading
support, and repair validation. Existing tree differencing is too strict, and
single-run dynamic comparison is unsound for concurrent, random, and
timing-dependent behavior.

The key observation is that equivalence for \scratchlang{} programs is
parametric in an \emph{observation lens}. We present a taxonomy that organizes
behavioral divergence by causal phenomenon and observing lens, and we build
\tool{}, an equivalence checker structured around that taxonomy. \tool{}
compiles projects into a causal intermediate representation of typed
resources and semantic transactions. It canonicalizes alpha-renamings, guards,
and procedure bodies; quotients same-trigger concurrency by Mazurkiewicz trace
normal forms over a conservative independence relation; separates program
order from races; and supports residual-frontier handling through SMT
obligations and counterexample-guided execution on the instrumented
\scratchlang{} VM. Every conclusive verdict carries explicit evidence:
equivalence by a bijection and trace quotient, difference by a typed witness,
and unresolved cases remain explicit unknowns.

We evaluate \tool{} on a fully automated, VM-witnessed mutation corpus built
from real \scratchlang{} projects, with per-lens ground-truth labels and a
deep-composition stratum that hides each defect under a stack of
equivalence-preserving refactorings. Under strict scoring that counts
abstention as error, \tool{} decides all \MatrixUsablePairs{} validated pairs
and makes $0/158$ false-equivalence claims on witnessed-different pairs;
structural, dynamic-only, and
large-language-model baselines fail on the classes predicted by the taxonomy,
ablations quantify the contribution of partial-order reduction and lens
parametricity, and the ambiguous-mutant study shows that targeted scenarios
expose divergences that random testing misses.
\end{abstract}

\section{Introduction}
\label{sec:intro}

\scratchlang{} is a large, real software ecosystem for introductory
programming. More than 140 million children have created over a billion
projects in it~\cite{scratch2024annual}. At this scale,
automated feedback, grading support, repair validation, and hint generation
all face the same question: when should two student programs be treated as
behaviorally equivalent? The question is hard because \scratchlang{} programs are
small event-driven systems: sprites, clones, variables, lists, monitors,
broadcasts, ask queues, pen and sound effects, and renderer-visible state run
on a cooperative green-thread VM~\cite{scratchvm}. The same behavior admits
many syntactic realizations (renames, script splitting, procedures,
additive-update rewrites), while a one-block edit can change behavior through
scheduling, random tokens, or frame timing; dropping
\texttt{broadcast and wait}, for example, removes a join edge whose absence
may appear only when a receiver is slow.

This setting makes \scratchlang{} a compact stress test for software
engineering methods. The language and environment were designed for broad
creative programming~\cite{resnick2009scratch,maloney2010scratch}, and
repository studies show that learners solve similar tasks through diverse
block structures, sprites, and control decompositions~\cite{aivaloglou2016kids}.
An equivalence checker used by graders, feedback systems, or repair validators
needs to preserve that diversity. It should accept correct refactorings across
solution styles and reject causal changes that alter what a learner, tutor, or
grader can observe.

Program comparison tools need to control false alarms on refactorings and
false equivalence claims on real bugs. Tree differencing~\cite{falleri2014gumtree,
zhang1989ted} over-approximates syntax because script boundaries and block
order are not semantic; single-run dynamic comparison under-approximates
behavior because it samples one schedule, random stream, and input
trace~\cite{stahlbauer2019testing}. In the debugging tutor that motivates
\tool{}, false equivalence is especially costly: it ends a learner's
debugging episode with the defect still present.

Event-driven visual programs require lens-parametric equivalence. A glide
and a jump can agree on final state and disagree on frames; two
initialization scripts can agree under one schedule and diverge under
another. The observer is part of the claim: final state, frame trace, monitors,
stage output, event causality, or debug trace.

We proceed in three steps. First, we define equivalence under an explicit
lens and a taxonomy linking causal mechanisms (state, event causality,
scheduling, randomness, time, clones, queues, persistent output), observation
lenses, and typed root causes (Section~\ref{sec:taxonomy}). The same
vocabulary drives verdicts, mutation operators, and per-lens labels.

Second, we build \tool{} around that taxonomy
(Sections~\ref{sec:model}--\ref{sec:algorithm}). Projects compile to CSIR:
typed resources, footprints (reads, writes, consumes, spawns, joins, kills),
and transactions cut at scheduler and lens-visible boundaries. Comparison
combines rename-invariant canonicalization (usage profiles, one
Weisfeiler--Leman refinement, bounded bijection search, expression and
procedure normal forms, control-region paths, dead-branch pruning) with a
Mazurkiewicz trace normal form~\cite{mazurkiewicz1987trace} over a
conservative independence relation. Residual mismatches become typed causal
differences, SMT side conditions, and targeted CEGAR scenarios for unresolved
frontiers on an instrumented \scratchlang{} VM~\cite{demoura2008z3,clarke2003cegar}.
The invariant is simple: every input that can influence lens-visible behavior
is compiled and compared.

Third, we evaluate with mechanical labels (Section~\ref{sec:evaluation}):
equivalence labels come from correct-by-construction transforms plus VM
falsification, difference labels require concrete VM witnesses, and mutants
that the random validation battery leaves unresolved form an explicit
ambiguous bucket. Baselines span
structural, opcode-abstraction, dynamic-only, and LLM judges.

We make the following contributions.
\begin{itemize}
\item A two-axis taxonomy of behavioral divergence in \scratchlang{} programs,
relating causal phenomena, observation lenses, and
typed root causes (Section~\ref{sec:taxonomy}).
\item Lens-parametric equivalence over CSIR, a causal intermediate
representation whose transactions carry typed footprints, scheduler cuts, and
provenance (Sections~\ref{sec:model}, \ref{sec:csir}).
\item A comparison algorithm that combines rename-invariant canonicalization,
trace normal forms with race structure, typed root-cause classification,
bounded bijection fallback for tied profiles and explanations, and
residual-frontier mechanisms based on SMT side conditions and VM-backed CEGAR
(Section~\ref{sec:algorithm}).
\item A reproducible, fully automated evaluation methodology for semantic
equivalence checkers: a VM-witnessed mutation corpus over real projects with
per-lens labels and a deep-composition stratum that separates semantic
reasoning from textual difference-spotting, released as a benchmark
together with operator manifests and generation scripts
(Section~\ref{sec:evaluation}).
\item An implementation and evaluation under strict scoring against
structural, dynamic, and LLM baselines, with ablations for partial-order
reduction and lens collapse.
\end{itemize}

\section{Running Example}
\label{sec:motivation}

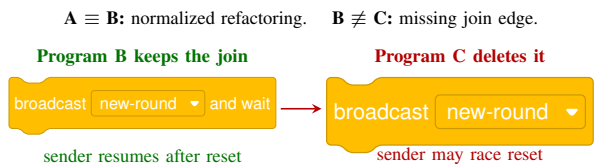
\begin{figure}[H]
\centering
\scriptsize
\begin{tikzpicture}[x=1cm,y=1cm,>=stealth]
  \tikzset{
    note/.style={font=\scriptsize, align=center, inner sep=1pt},
    headline/.style={font=\scriptsize, align=center, text width=.99\columnwidth},
    bad/.style={draw=red!70!black, line width=.60pt, ->}
  }
  \node[headline] at (3.45,.20) {
    \textbf{A $\equiv$ B:} normalized refactoring.
    \quad
    \textbf{B $\not\equiv$ C:} missing join edge.
  };
  \node[note,text=green!45!black] at (1.35,-.30)
    {\textbf{Program B keeps the join}};
  \node[note,text=red!70!black] at (5.55,-.30)
    {\textbf{Program C deletes it}};
  \node[anchor=north] (blockB) at (1.35,-.47) {%
    \resizebox{.40\columnwidth}{!}{%
    \begin{scratch}
    \blockevent{broadcast \selectmenu{new-round} and wait}
    \end{scratch}}};
  \node[anchor=north] (blockC) at (5.55,-.47) {%
    \resizebox{.40\columnwidth}{!}{%
    \begin{scratch}
    \blockevent{broadcast \selectmenu{new-round}}
    \end{scratch}}};
  \draw[bad] (3.18,-.98) -- (3.72,-.98);
  \node[note,text=green!45!black] at (1.35,-1.60) {sender resumes after reset};
  \node[note,text=red!70!black] at (5.55,-1.60) {sender may race reset};
\end{tikzpicture}
\caption{Running example. A distant refactoring (A/B) is equivalent, while a
one-block change (B/C) removes the broadcast join edge and becomes visible
under event-causal and frame-observing lenses.}
\label{fig:running-example}
\Description{A compact Scratch-style comparison showing equivalent refactoring
and a one-block broadcast difference.}
\end{figure}

Figure~\ref{fig:running-example} distills the comparison problem that recurs
throughout the paper. Program A is a reference game fragment: a bowl
initializes score and lives, asks the apple to reset, and increments score
when the bowl touches the apple. The apple reset handler hides the sprite,
moves it to a random top-row position, and shows it again.

Program B looks very different from Program A: it renames variables, sprites,
and messages; splits initialization into two green-flag scripts; extracts the
reset logic into a custom block; and rewrites \texttt{change score by 1} as an
assignment. A tree differencer sees many edits. The programs are behaviorally
equivalent: resources admit a bijection, the initialization scripts
write disjoint variables and commute, the custom block has the same canonical
body as the original receiver, and the additive update normalizes to the same
transaction.

Program C changes only one block and diverges from Program B. The nonblocking
broadcast removes the join edge from receiver completion to sender
continuation. If fruit reset takes a frame, or if the next collision arrives
before the receiver finishes, the bowl can continue while the fruit still has
its old state. Final score may coincide, while event causality, frame trace,
or stage-visible behavior differs. The verdict names the observation lens:
A versus B is accepted with an explicit bijection and trace quotient; B versus
C is rejected under lenses that observe the missing join edge, with a
\textsc{MissingJoinEdge} witness.
The example also fixes the role of the main abstractions used below. Variables,
sprites, and messages become typed CSIR resources; independent initialization
writes commute only when footprints prove independence; and the removed wait is
an event-causal fact whose visibility depends on the selected lens.

\section{A Taxonomy of Behavioral Divergence}
\label{sec:taxonomy}

The taxonomy asks, for each behavioral divergence, what mechanism causes it,
which lenses observe it, and which typed root cause a checker should report.
We derived it from the VM's semantic carriers (Section~\ref{sec:model}) and
cross-checked it against LitterBox bug patterns~\cite{fraser2021litterbox,
fraedrich2020bugs}.

\subsection{Axis 1: Observation Lenses}

A lens is a projection from concrete traces to observations.
Table~\ref{tab:lens} lists the lenses \tool{} supports. \ldefault{} is the
union of frame-visible, stage-visible, monitor, and event-causal
observations; it corresponds to what an attentive user of the running project
can perceive. Lenses form a partial order when one projection includes
another's observations. We use ``stronger'' only for comparable lenses; for
incomparable projections, \tool{} reports a verdict vector indexed by lens.

\begin{table}[H]
\centering
\caption{Observation lenses. Each preserves a subset of trace information;
\ldefault{} combines the middle four.}
\label{tab:lens}
\begin{tabular}{ll}
\toprule
\textbf{Lens} & \textbf{Preserved observations}\\
\midrule
\lfinal{} & Final variables, lists, target state, clone count\\
\lframe{} & Frame-boundary visual snapshots, yield structure\\
$L_{\mathit{stage}}$ & Sprite, backdrop, pen, speech, sound effects\\
\lmonitor{} & Monitor values, visibility\\
\levent{} & Broadcast, join, ask, clone, stop causality\\
$L_{\mathit{debug}}$ & Primitive-level VM trace\\
\bottomrule
\end{tabular}
\end{table}

\subsection{Axis 2: Causal Phenomena}

We organize supported \scratchlang{} divergences into eight carrier families.
Each row of Table~\ref{tab:taxonomy} names the phenomenon, the weakest lens
family under which it becomes observable, the typed root causes \tool{}
reports for it, and the semantic carrier that CSIR must preserve. Operator
manifests in the artifact map these rows to generated mutations
(Section~\ref{sec:corpus}).

\begin{table*}[t]
\centering
\caption{Taxonomy of behavioral divergence. ``Lens floor'' is the weakest
lens family that observes the phenomenon; comparable stronger lenses inherit
the difference.}
\label{tab:taxonomy}
\small
\setlength{\tabcolsep}{4pt}
\begin{tabular}{@{}>{\raggedright\arraybackslash}p{0.19\textwidth}>{\raggedright\arraybackslash}p{0.24\textwidth}>{\raggedright\arraybackslash}p{0.11\textwidth}>{\raggedright\arraybackslash}p{0.36\textwidth}@{}}
\toprule
\textbf{Phenomenon} & \textbf{Carrier} & \textbf{Lens floor} & \textbf{Typed root causes}\\
\midrule
P1 State transformation & variables, lists, properties & \lfinal{} & ValueChange; GuardChange; UninitializedRead\\
P2 Event causality & broadcasts, joins, stops, triggers & \levent{} & Missing/ExtraJoinEdge; BroadcastEdgeRemoved; TriggerChange; Missing/ExtraKillEdge\\
P3 Schedule sensitivity & same-trigger thread order & \levent{}/\lfinal{} & race-structure mismatch\\
P4 Randomness, environment & random stream, input tokens & any & RandomStreamShift\\
P5 Time and frames & yields, waits, glides, timer & \lframe{} & ChangedFrameBoundary; FramePathChange; ChangedTimerCausality\\
P6 Clone lifecycle & clone families, clone-start hats & \levent{}/$L_{\mathit{stage}}$ & ChangedCloneMultiplicity; CloneInitChange\\
P7 Interaction queues & ask FIFO, answer & \levent{} & AskQueueOrderChanged\\
P8 Persistent output & pen buffer, monitors, sound, looks & $L_{\mathit{stage}}$/\lmonitor{} & PenEffectChange; EffectRemoved/Added; MonitorVisibleOnly\\
\bottomrule
\end{tabular}
\end{table*}

Three properties matter downstream. Root causes are typed by carrier
resource, so \textsc{MissingJoinEdge} names a broadcast message and
\textsc{UninitializedRead} names a variable. Lens floors make labels
vectors: a P5 operator can be different under \ldefault{} and equivalent
under \lfinal{}. Finally, each phenomenon maps to CSIR footprint facts,
letting the classifier recover root causes without syntax patterns and
letting per-root-cause results double as per-phenomenon coverage.
Two cases illustrate the boundary: same-trigger scripts that both write one
variable become a P3 race fact, and concurrent ask blocks can reorder the
global FIFO prompt queue under P7.

The taxonomy also acts as a soundness contract. Legal simplifications
preserve the carrier and the weakest observing lens. Renaming a variable
preserves P1 facts only if all reads, writes, and monitor references move
together. Reordering two scripts is valid only when their footprints prove
independence; when both scripts write the same resource, the comparison
records a P3 race fact. Likewise, replacing a glide by a move is an
equivalence under \lfinal{} and a difference under \lframe{}. The contract
prevents a single binary verdict from hiding why two reasonable observers
can disagree.

\section{Semantic Model}
\label{sec:model}

\subsection{Concrete Configurations}

We model a \scratchlang{} project as a transition system over configurations
\[
C = (\Sigma, \Theta, H, A, K, E, F, O),
\]
where $\Sigma$ is the store and visual state; $\Theta$ the green-thread pool;
$H$ hats and pending events; $A$ the ask queue and answer; $K$ clone families;
$E$ environment and random tokens; $F$ frame, time, and redraw state; and $O$
observations. The model follows the VM implementation~\cite{scratchvm}. Four
facts drive the IR and corpus: threads run cooperatively within frames; loops
yield once per iteration, so bounded unrolling must preserve synthetic frame
barriers under \lframe{}; broadcasts use casefolded message identity and
restart running receivers; and ask-and-wait serializes through one global FIFO
question queue.

\subsection{Lens-Parametric Equivalence}

A lens $L$ projects concrete traces to observations (Table~\ref{tab:lens}).
Two projects are equivalent under $L$ when every admissible environment,
random, time, and input stream produces equal $L$-projected traces, modulo
alpha-renaming, stuttering of $L$-unobservable transactions, reordering of
independent same-trigger transactions, and permutation of indistinguishable
clone families. The schedule quotient treats the relative order of distinct
same-trigger scripts as nondeterministic and reports dependence on it as a
race (Section~\ref{sec:algorithm}).

The quotient preserves each lens-visible cut. Stuttering erases only
$L$-unobservable transactions; broadcast joins remain visible under
\levent{}, redraw boundaries under \lframe{}, and monitor toggles under
\lmonitor{}. The alpha-renaming component is typed as well: a stage
variable, a sprite-local variable, a list region, and a broadcast message
live in different resource spaces with separate maps. The restrictions let
the implementation use compact canonical products while retaining a direct
connection to VM behavior.

\section{Causal \scratchlang{} IR}
\label{sec:csir}

\subsection{Typed Resources and Footprints}

{\sloppy
CSIR types every semantic carrier:
\textsc{Var}, \textsc{List}, \textsc{TargetProp}, \textsc{Monitor},
\textsc{AskQueue}, \textsc{Answer}, \textsc{RandomStream}, \textsc{Timer},
\textsc{BroadcastMessage}, \textsc{CloneFamily}, \textsc{ThreadSet},
\textsc{FrameBoundary}, \textsc{PenBuffer}, \textsc{SoundChannel},
\textsc{EnvToken}, \textsc{External}. Resources carry scope (global,
sprite-local, clone-local) and, for lists, a region (whole, length, index,
suffix). A footprint records reads, writes, creates, deletes, ordered-token
consumes, observes, spawns, joins, kills, frame barriers, and commutative
effects such as additive variable updates.\par}

\subsection{Transactions and Compilation}

A transaction is a maximal effect sequence between scheduler- or
lens-visible cuts: hats, waits, ask-and-wait, broadcasts and joins, clone
birth and death, stops, loop back-edges, promise waits, frame boundaries, and
monitor updates. The compiler enforces the invariant that every
behaviorally relevant input is compiled: guards and reporter arguments carry
read/token footprints; menus resolve to selections; inline variable
references become reads; bounded literal loops unroll with one synthetic
frame barrier per iteration; statically false branches are pruned; sound
procedures are inlined and the rest compared by canonical content digest.
Unsupported extensions compile to opaque external resources; they reduce
decisiveness and produce unknowns when their effects matter.

Each compiled transaction records its trigger, target family, lexical
provenance, normalized primitive sequence, control-region path, footprint,
and $L$-visible observations. The product compared by the algorithm is the
multiset of these facts after alpha canonicalization and trace
normalization. Facts are intentionally redundant: a broadcast contributes a
message resource, a spawn edge, a receiver set, and optionally a join edge.
The redundancy makes root-cause classification local. Removing
\texttt{broadcast and wait}, for instance, changes a join fact without
requiring the classifier to reconstruct a dynamic happens-before graph from
raw blocks.

\section{Comparison Algorithm}
\label{sec:algorithm}

Algorithm~\ref{alg:kernel} summarizes the comparison kernel. Every accepting
path carries a witness: equality of canonical products, an explicit alpha
bijection, or equality of the \lfinal{} abstract transfer. Mismatches become
typed root causes plus obligations; targeted VM scenarios exercise only
residual frontiers, and unproved cases remain explicit unknowns.
Obligations are closed side conditions with a status in
\{\emph{proved}, \emph{refuted}\}. Open predicates are recorded as
frontiers.

\begin{algorithm}[t]
\caption{\tool{} lens-parametric product comparison}
\label{alg:kernel}
\small
\begin{algorithmic}[1]
\Require projects $P_r,P_s$; lens $L$; optional VM budget $B$
\State $C_r,C_s \gets \Call{CompileCSIR}{P_r,P_s}$
\State $M_r,M_s \gets \Call{AlphaCanon}{C_r,C_s}$ \Comment{usage profiles + WL}
\State $(F_r,O_r) \gets \Call{ProductFeatures}{C_r,L,M_r}$
\State $(F_s,O_s) \gets \Call{ProductFeatures}{C_s,L,M_s}$
\If{$F_r = F_s$ and \Call{Closed}{$O_r \cup O_s$}}
  \Return \textsc{Equivalent} with $(M_r,M_s)$ and certificates $O_r \cup O_s$
\ElsIf{$F_r = F_s$}
  \Return \textsc{Unknown} with frontier \Call{Open}{$O_r \cup O_s$}
\EndIf
\State $(F_s',\pi,\delta,O_\pi) \gets \Call{BoundedTieSearch}{F_r,F_s,C_s,M_s}$
\If{$\delta = 0$ and \Call{Closed}{$O_r \cup O_\pi$}}
  \Return \textsc{Equivalent} with explicit bijection $\pi$
\ElsIf{$\delta = 0$}
  \Return \textsc{Unknown} with frontier \Call{Open}{$O_r \cup O_\pi$}
\EndIf
\State $(T_r,T_s,O_t) \gets \Call{FinalTransfer}{C_r,C_s,L}$
\If{$L=\lfinal{}$ and $T_r=T_s$ and \Call{Closed}{$O_t$}}
  \Return \textsc{Equivalent under } \lfinal{} with certificates $O_t$
\ElsIf{$L=\lfinal{}$ and $T_r=T_s$}
  \Return \textsc{Unknown} with frontier \Call{Open}{$O_t$}
\EndIf
\State $D \gets \Call{Classify}{C_r,C_s,F_r,F_s',L}$ \Comment{typed root causes}
\State $U \gets \Call{Obligations}{D}$ \Comment{solver fragment or frontier}
\If{$B>0$ and $U$ contains a conditional frontier}
  \State $S \gets \Call{Scenarios}{U}$
  \State $W \gets \Call{RunTargetedVM}{P_r,P_s,S,B}$
  \If{$W$ confirms divergence}
    \State \Return \textsc{Different}$(W)$
  \EndIf
\EndIf
\If{\Call{LensSufficient}{$D,U,L$}}
  \State \Return \textsc{Different}$(D,U)$
\EndIf
\State \Return \textsc{Unknown} with frontier $U$
\end{algorithmic}
\end{algorithm}

The last conclusive branch is deliberately narrow:
\Call{LensSufficient}{$D,U,L$} holds only for footprint-backed root causes
whose observability under $L$ is independent of unresolved predicates.
Masked-comparison items, unsupported reporters, and any root cause whose
observability depends on an open obligation remain in $U$ until solver
discharge or a VM witness supports a conclusive \textsc{Different} verdict.

\subsection{Canonicalization}

\tool{} canonicalizes both projects independently, then compares canonical
feature multisets. Expression trees fold constants, orient comparisons, sort
commutative operands, simplify negations, and normalize
\texttt{change v by c} with \texttt{set v to v + c}. Every primitive also
carries a \emph{control-region path}: enclosing control opcodes, branch
indices, and guard expressions. Thus \texttt{if c \{set a; set b\}} and
\texttt{if c \{set a\}; set b} no longer flatten to the same product; the
former tags both writes with \texttt{if\#b0[c]}, the latter only the write to
\texttt{a}. Statically decided guards contribute no path element.

Resources receive canonical indices per kind and scope, ordered by a
rename-invariant usage profile refined by one Weisfeiler--Leman round over
co-footprint neighborhoods~\cite{weisfeiler1968reduction}. If the base
comparison fails, \tool{} searches remaining bijections over profile-tied
resources by coordinate descent and a bounded small-space sweep. An
equalizing bijection is a sound equivalence witness; otherwise the best
bijection only minimizes the explanation delta.

Usage profiles use semantic roles. A variable written by a green-flag
initialization, read by a guard, and shown as a monitor has a different
profile from a variable written only inside a clone-start script. The WL
refinement adds neighborhood context: two variables that both appear in
arithmetic updates can still separate if one co-occurs with a broadcast join
and the other with a renderer-visible motion effect. Remaining ties are
common in small projects with symmetric sprites or duplicate counters; the
bounded search closes those cases inside proof construction.

\begin{algorithm}[t]
\caption{Bounded bijection search over profile-tied resources}
\label{alg:tiesearch}
\small
\begin{algorithmic}[1]
\Require canonical features $F_r$, $F_s$; tie groups $G_1,\dots,G_k$
\State $F_{best} \gets F_s$;\quad $\pi_{best} \gets \emptyset$;\quad $O_{best}\gets\emptyset$;\quad $\delta^* \gets |F_r \ominus F_s|$
\For{$i \gets 1$ \textbf{to} $k$} \Comment{coordinate descent, cost $\sum_i |G_i|!$}
  \If{no canonical token of $G_i$ occurs in $F_r \ominus F_{best}$} \textbf{continue} \EndIf
  \For{permutation $\pi$ of $G_i$, others fixed at $\pi_{best}$}
    \State recompute features and certificates $(F_\pi,O_\pi)$;\; $\delta \gets |F_r \ominus F_\pi|$
    \If{$\delta = 0$} \Return $(F_\pi,\pi,0,O_\pi)$ \EndIf
    \If{$\delta < \delta^*$} $F_{best} \gets F_\pi$;\; $\pi_{best} \gets \pi$;\; $O_{best}\gets O_\pi$;\; $\delta^* \gets \delta$ \EndIf
  \EndFor
\EndFor
\State full cross-product sweep when $\prod_i |G_i|! \le 64$, budget $1024$
\State \Return $(F_{best},\pi_{best},\delta^*,O_{best})$ \Comment{minimal-delta bijection}
\end{algorithmic}
\end{algorithm}

Algorithm~\ref{alg:tiesearch} gives the search. Coordinate descent costs
$\sum_i |G_i|!$; the relevance filter skips groups absent from the residual
delta, and the full cross-product appears only under the small-space budget.
In the corpus every proven equivalence closes by direct canonical equality;
the search mainly provides a bounded fallback and explanation minimizer.

Procedures are keyed by defining sprite. Bodies are compared by content
digest over alpha-canonicalized primitives in the same trace normal form as
transaction clusters (Section~\ref{sec:tracenf}). The digest triple--raw,
guard-masked, and constant-masked--lets the classifier distinguish
\textsc{GuardChange} and \textsc{ValueChange} inside procedures and keeps
procedure mismatches typed. Digests reach a call-graph fixpoint, accepting
renamed-identical procedures and rejecting same-named different ones.

\subsection{Trace Normal Forms with Race Structure}
\label{sec:tracenf}

Within each trigger cluster, \tool{} proves pairwise independence from typed
footprints. Transactions are independent only when read/write conflicts are
absent (modulo declared commutative effects), neither consumes ordered
tokens, $L$-visible observations commute, and no spawn, join, kill, or frame
barrier fixes order. Unproved pairs are dependent. Within one script,
dependence becomes program order; across scripts it becomes a race fact.
Each cluster maps to the lexicographically least linearization of this
partial order, a Mazurkiewicz trace normal form~\cite{mazurkiewicz1987trace,
godefroid1996partial}: interleavings of the same trace match, ordered scripts
never match races, and fully independent transactions compare as a multiset.

For example, two green-flag scripts that both write \texttt{score} emit an
unordered race fact. Merging them into one ordered script changes the product;
renaming the variable preserves the race fact under the alpha map.
Equivalence preserves scheduler commitments as well as effects.

\begin{algorithm}[t]
\caption{Trace normal form for one trigger cluster}
\label{alg:tracenf}
\small
\begin{algorithmic}[1]
\Require transactions $T$ with footprints, program-order edges $E_p$, lens $L$
\State $E \gets E_p$;\quad $R \gets \emptyset$
\ForAll{unordered pairs $(t_i,t_j) \in T$}
  \If{\Call{Independent}{$t_i,t_j,L$}}
    \State continue
  \EndIf
  \If{$t_i$ and $t_j$ come from distinct same-trigger scripts}
    \State $R \gets R \cup \{\Call{RaceFact}{t_i,t_j}\}$
  \Else
    \State $E \gets E \cup \{\Call{ProgramOrder}{t_i,t_j}\}$
  \EndIf
\EndFor
\State $N \gets$ lexicographically least topological order of $(T,E)$
\State $F \gets \Call{EmitFacts}{N,L} \cup R$
\State \Return $(F,R)$
\end{algorithmic}
\end{algorithm}

Algorithm~\ref{alg:tracenf} is the step that keeps commutation and races
separate. Independence is checked on typed footprints, ordered-token
consumption, lens-visible observations, and lifecycle edges. A dependent pair
inside one script becomes program order; a dependent pair across same-trigger
scripts becomes an unordered race fact. The canonical product then contains
both the least linearization and the race facts. The representation accepts
script splitting and commuting initializers while preserving scheduler
commitments that a learner or tutor can observe under \levent{} or
\lframe{}.

\subsection{Verdicts, Classification, and Obligations}

Equal canonical products prove equivalence only when their side conditions are
closed. Under \lfinal{}, an abstract transfer over the numeric, string, list,
clone, and monitor domains provides a second sound acceptance path when
products differ only by final-state-invisible structure and the transfer
obligation is proved. Unequal products enter the typed
classifier, which recovers the taxonomy's root causes from canonical fact
deltas: join and spawn facts for P2, trigger sets for trigger changes,
dangling send sets for broken broadcast wiring, read-before-write summaries
for uninitialized reads, clone-trigger feature isolation for clone
initialization, pen and observation totals for P8, and masked feature
comparison for guard and literal changes (a pair whose features equalize
when guards are masked differs exactly in its guards). Classified items
carry confidence. Footprint-backed kinds are sound: a missing join fact is
a fact about compiled structure, and no execution can retract it.
Masked-comparison kinds are conditional: guard-masked equality isolates the
change to guard text, and semantic equivalence of those guards is a
satisfiability question. The item becomes an obligation and, when the solver
fragment leaves it open, a CEGAR frontier with a synthesized scenario.
Supported obligations, including guard feasibility, list abstraction
equality, clone partition equality, and broadcast join equivalence, discharge
through Z3 when available and through an exact finite fallback
otherwise~\cite{demoura2008z3}. Discharged obligations become certificates;
open obligations stay in the frontier and are excluded from equivalence and
static \textsc{Different} certificates.

\subsection{CEGAR with the Instrumented VM}

Conditional frontiers become targeted scenarios: a random-stream frontier
yields a scenario with two distinguishable token prefixes whose constraints
($r_i \neq r_j$, range bounds) the solver instantiates; an ask-order
frontier yields distinguishable answers; a join frontier yields a
receiver-stress run. The runner executes both projects under identical
deterministic oracles, records primitive, event, clone, monitor, and
renderer signals, aligns trace segments back to CSIR transactions, and
confirms or refutes the frontier. Spurious evidence mutates an explicit
refinement state (random taint to indexed tokens, clone families to
partitions, final summaries to frame snapshots, renderer hashes to pixel
regions) and re-enters comparison, in the counterexample-guided
tradition~\cite{clarke2003cegar}.

Frontiers are structured records. Each record names the carrier, lens,
trigger path, unresolved predicate, candidate scenario generator, and the
observations that would close the case. A random frontier records token
indices and value constraints; an ask frontier records question order and
answer substitutions; a renderer frontier records the region or target
property that must diverge. The record structure lets the implementation cache
negative runs, refine only the affected abstraction, and report
\textsc{Unknown} with enough context for a stronger scenario or human review.

\subsection{Soundness Boundary}

\tool{} is intentionally asymmetric: finite VM runs prove difference, while
equivalence requires an accepting static path. Table~\ref{tab:soundness}
summarizes the contract. The theorem scope is the supported Scratch-VM subset
compiled to CSIR: core variables/lists, target properties, broadcasts and
joins, green-thread yields, clones, ask queues, monitors, renderer-facing
state, and the reporters covered by the expression normalizer. Opaque
extensions, unsupported reporters, solver timeouts, and open frontiers yield
\textsc{Unknown} unless a concrete witness is found.

\begin{table}[t]
\centering
\caption{Verdict paths and guarantees.}
\label{tab:soundness}
\scriptsize
\setlength{\tabcolsep}{2.5pt}
\begin{tabular}{>{\raggedright\arraybackslash}p{0.25\columnwidth}>{\raggedright\arraybackslash}p{0.39\columnwidth}>{\raggedright\arraybackslash}p{0.25\columnwidth}}
\toprule
\textbf{Path} & \textbf{Guarantee} & \textbf{Known exclusions}\\
\midrule
Canonical product equality & Lens equivalence with closed side conditions, modulo alpha-renaming, POR, stuttering, and procedure digests & Opaque extensions; unsupported reporters\\
Tie-group bijection & Same guarantee after bounded bijection search with closed side conditions & Search may abstain\\
\lfinal{} transfer & Final-state equivalence over stated abstract domains with proved transfer obligation & No frame/event/stage guarantee\\
Static causal fact & Difference for the named lens-visible carrier when root cause is lens-sufficient & Other lenses may agree\\
VM witness & Concrete counterexample trace & No equivalence proof\\
\textsc{Unknown} & No semantic claim & Incomplete by design\\
\bottomrule
\end{tabular}
\end{table}

\subsection{End-to-End Verdict Paths}

The kernel has only a few accepting shapes, and those shapes cover the
cases that make Scratch comparison difficult. Consider an alpha-renamed
project whose initialization script is split into two green-flag scripts.
CSIR compilation assigns the renamed variables the same usage profiles,
the split creates two transactions under the same trigger, and the
independence relation proves that their footprints commute. The trace
normal form emits the same product facts on both sides: the
equivalence witness is the alpha map plus the trace quotient. Block-level
similarity has no role in this proof path.

For a missing \texttt{broadcast and wait}, compilation emits the same
message resource and receiver-spawn facts and omits the join fact from
receiver completion to sender continuation. The missing fact is a P2
mismatch, so the
classifier reports \textsc{MissingJoinEdge} and names the message and
receiver family. If the downstream observation is immediate, the static
root cause is enough; if the difference requires receiver contention, the
obligation synthesizes a receiver-stress scenario. A confirmed VM trace is
a concrete witness; an unconfirmed scenario leaves the pair conditional.

For a glide replaced by a jump, the products differ in frame-path facts and
renderer-visible target positions under \ldefault{} or \lframe{}. The same
pair can still be accepted under \lfinal{} if the abstract transfer proves
equal final coordinates, size, direction, variables, lists, and clone
counts. The example shows why the verdict carries a lens:
a single global label would either miss a visible frame difference or
incorrectly reject a valid final-state equivalence.

For guard and literal changes, the classifier first asks which masked
product equalizes the pair. If guard-masked products match and raw products
differ, the residual obligation is semantic guard equivalence. Simple numeric
and string fragments discharge
through the solver or finite fallback. If the guard lies outside the
supported fragment, the frontier records the unknown expression and the
transaction it controls; targeted execution may find a witness, and the
static checker keeps the claim conditional until then.

\section{Implementation}
\label{sec:impl}

\tool{} comprises a Python analysis kernel and a Node.js VM harness. The
kernel implements CSIR compilation, canonicalization, trace normal forms,
classification, obligations, and refinement. The harness wraps
\texttt{scratch-vm} v5.0.300 (repository SHA \texttt{e6f5711})~\cite{scratchvm}
with deterministic random, time, keyboard, broadcast, and ask oracles,
recording tick snapshots plus primitive, event, clone, monitor, renderer, and
collision signals; without headless GL it falls back to
bounding boxes while preserving non-pixel traces. The kernel decides nothing
by VM execution: VM runs provide witnesses and refutations only. The
implementation passes $189$ unit and soundness-regression tests covering the
hardening fixes found during corpus development, including guard capture,
menu resolution, loop-yield unrolling, sprite-scoped procedures,
control-region paths, masked procedure digests, race structure, and tie
search.
The review artifact releases a runnable core subset: CSIR compilation,
normalizers, canonical product comparison, trace normal forms, the static
verdict driver for the controlled suite and non-LLM baselines, the VM trace
harness, and aggregation scripts. It excludes private tutor integration,
service deployment code, and credentials.
The same harness validates labels, dynamic baselines, and targeted frontier
scenarios, so oracle regressions produce quarantine spikes before labels can
shift silently. Several regressions came directly from corpus
development: inline-reference decoding, boundary-aware canonicalization of
bare names, procedure-body normal forms, and coordinate-descent tie search
were all hardened after generated pairs exposed a failure mode.

\section{Evaluation}
\label{sec:evaluation}

The evaluation is designed to run without a human in the loop: every label
is either correct by construction or witnessed by VM execution, abstention
is a first-class outcome for every method, and false equivalence is the
headline error. It asks six research questions.

\begin{description}
\item[RQ1] How accurately does \tool{} classify equivalent and different
pairs on the labeled corpus, and how often does it abstain?
\item[RQ2] When pairs differ, does \tool{} recover the injected root cause?
\item[RQ3] How does \tool{} compare with structural, abstraction,
dynamic-only, and LLM baselines, in accuracy and in false-equivalence count?
\item[RQ4] On mutants left unresolved by a random scenario battery, how often
do targeted scenarios expose the divergence?
\item[RQ5] What do partial-order reduction and lens parametricity
contribute, and do per-lens verdicts match per-lens labels?
\item[RQ6] What does the pipeline cost, and how often does each escalation
layer decide?
\end{description}

\subsection{Corpus Construction}
\label{sec:corpus}

Seeds are real \scratchlang{} projects that parse, compile, and execute
headlessly, deduplicated by opcode multiset and stratified by size. Of $94$
scanned projects, $60$ are accepted ($8$ too small, $26$ above the block cap),
spanning $5$--$600$ blocks; $44$ use broadcasts, $18$ clones, $17$ randomness,
$6$ ask-and-wait, and $4$ extension blocks. Each seed yields identity pairs,
semantics-preserving transforms, lens-differential transforms, and
behavior-changing mutants instantiating Table~\ref{tab:taxonomy}; some
different mutants are further wrapped in benign refactoring noise.
Operators mutate only code reachable from real event hats.

The operator families are paired to the taxonomy. Equivalence operators
rename variables and messages, split independent scripts, insert statically
dead branches outside loop bodies, extract procedures, and rewrite additive
updates. Lens-differential operators preserve final state while changing an
observable path, for example replacing a glide with a jump, inserting a
wait, or changing loop-yield structure. Different operators remove or add
joins, stops, triggers, guards, parameters, clone creation, message wiring,
or observable effects. This design gives each admitted different pair both
a semantic label and an injected root cause, while keeping equivalence pairs
rich enough to defeat textual comparison.

Labels are mechanical. Equivalent-labeled pairs face a deterministic
falsification battery whose stimuli are translated through the recorded
identifier bijection and whose observations are translated back; a renamed
message is the same logical broadcast on both sides. Reproducible
divergences quarantine the pair, while isolated \lfinal{} failures downgrade
only that lens claim. Different-labeled pairs require a concrete VM witness;
unwitnessed mutants enter the ambiguous bucket for RQ4. Labeling,
dynamic-baseline, and \tool{} CEGAR seeds are disjoint. The shipped corpus has
\MatrixTotalPairs{} generated pairs: \MatrixUsablePairs{} headline pairs,
\MatrixAmbiguous{} ambiguous pairs, and \MatrixQuarantined{} quarantined
pairs ($2.2\%$). The quarantine history found computed-message renames,
loop-body dead branches, reporter deletions, and two oracle-stimulus defects;
each became a precondition or regression.
The battery spans green-flag smoke runs, per-key taps and sweeps, sprite and
stage clicks, broadcast injections and bursts, ask-and-answer scripts, and
random-stream fuzzing under deterministic seeds. A pair whose full trace
diverges is quarantined only when the divergence reproduces under the same
seed, filtering borderline-tick flakes. This validation policy is asymmetric
by construction: equivalence-labeled pairs survive unless falsified, and
different-labeled pairs require a positive witness.

\begin{table}[t]
\centering
\caption{Corpus composition by operator: pairs generated and pairs
admitted to the headline metrics after validation. Different-labeled pairs
need a VM witness, so rare-schedule operators have lower admission rates.}
\label{tab:operators}
\small
\resizebox{\columnwidth}{!}{%
\begin{tabular}{llrr}
\toprule
\textbf{Operator} & \textbf{Class} & \textbf{Gen.} & \textbf{Usable}\\
\midrule
identity & control & 60 & 60\\
\texttt{eq\_alpha} & EQ & 37 & 36\\
\texttt{eq\_dead\_branch} & EQ & 35 & 34\\
\texttt{eq\_msg\_rename} & EQ & 26 & 24\\
\texttt{eq\_add\_xform} & EQ & 15 & 15\\
\texttt{eq\_commute} & EQ & 5 & 5\\
\texttt{ld\_wait\_add} / \texttt{glide} / \texttt{unroll} & lens-diff & 34 & 18\\
\texttt{df\_event\_swap} & DF & 25 & 14\\
\texttt{df\_block\_del} & DF & 25 & 12\\
\texttt{df\_stop\_add} & DF & 20 & 17\\
\texttt{df\_join\_add} / \texttt{drop} & DF & 21 & 10\\
\texttt{df\_param} & DF & 18 & 12\\
\texttt{df\_msg\_break} & DF & 16 & 6\\
\texttt{df\_cond\_flip} & DF & 16 & 12\\
\texttt{df\_clone\_dup} / \texttt{init\_del} & DF & 11 & 6\\
hard\_eq (4--6 EQ ops) & composed & 120 & 112\\
hard\_df (1 DF + 3--5 EQ) & composed & 60 & 51\\
\midrule
Total & & 544 & 444\\
\bottomrule
\end{tabular}
}%
\end{table}

Admission is asymmetric: equivalence pairs survive VM falsification, and
different pairs require positive witnesses. Rare-schedule operators
populate the ambiguous bucket. The composed stratum stacks four to
six equivalence transforms, or hides one bug under three to five such
transforms, separating semantic reasoning from textual matching.
The stratum models the tutoring case in which a learner both refactors and
introduces one defect: the textual diff is large, and the semantic
difference should remain a single root cause. Rendered-text judges get a
shortcut on single edits; composition removes it.

\subsection{Methods Under Comparison}

\tool{} runs in four configurations: full (static kernel with CEGAR),
static-only, without partial-order reduction, and under \lfinal{} only.
Static-only reports the kernel verdict before frontier escalation; on the
headline corpus it coincides with full \tool{} because no headline verdict
requires escalation. The
non-LLM baselines represent three method families: canonical fingerprint
equality and normalized tree-edit distance (structural); opcode multiset and
ordered opcode traces (abstraction); and dynamic-only comparison, which runs
$N{=}5$ deterministic scenarios and declares equivalence when no
divergence appears. The dynamic baseline compares traces under the
same identifier-translated quotient oracle as label validation, so its
errors reflect genuine sampling brittleness, with identifier artifacts
removed by the shared quotient.

The LLM tier evaluates GLM-5.1, Qwen3.6-plus, and Kimi~K2.6 through one
OpenAI-compatible gateway, with served model identifiers, prompts, raw
responses, and parsed JSON pinned in the artifact. Each model sees
scratchblocks text, the \ldefault{}
definition, and the closed taxonomy, then returns strict JSON with verdict,
optional root cause, and confidence; \textsc{Unknown} is available, and
persistent malformation counts as abstention.
Requests use per-pair on-disk checkpoints and a single retry for malformed
JSON, preventing transient gateway failures from re-judging decided pairs. The
models do not see labels, VM witnesses, \tool{} verdicts, or another model's
output.

All methods see the same pairs and no labels, witnesses, or other outputs.
Strict overall accuracy counts abstention as wrong; decided accuracy and
coverage separate selectivity. Diagnosis requires both verdict and injected
root cause. We also report different-class precision/recall, false
equivalent/different counts, bootstrap $95\%$ intervals, McNemar tests, and
single-edit versus deep-composition strata~\cite{mcnemar1947note,
efron1994bootstrap}.

Strict scoring matches deployment. A selective judge that returns
\textsc{Unknown} on hard equivalences has no basis for suppressing false
alarms as a tutor gate. We report both the selective and deployment views,
and we isolate false equivalence because that error certifies a defective
program as correct. False differences remain undesirable; in the motivating
workflow they produce recoverable hints or review requests.

\subsection{Experimental Environment}

Experiments run on one 16-core Windows workstation. The VM stages perform
about $19{,}000$ deterministic \texttt{scratch-vm} executions; the LLM tier
issues about $1{,}900$ checkpointed gateway requests, under \$10 total.
The artifact records raw per-pair outputs, seeds, scenarios, and model
responses before aggregation. The non-LLM matrix can be regenerated from
the shipped corpus without network access; the LLM tables can be re-scored
from pinned raw responses even if served model versions drift.

\subsection{RQ1--RQ2: Effectiveness}

Table~\ref{tab:matrix} reports \MatrixUsablePairs{} validated pairs. \tool{}
decides every pair with accuracy \MatrixSpectraAccuracy{}, different-class
precision/recall \MatrixSpectraPrecision{}/\MatrixSpectraRecall{}, and
\MatrixSpectraFalseEq{} false equivalences, at a $0.22$s median. The
false-equivalence count is $0/158$ on the validated different class; by the
rule of three, the corresponding upper $95\%$ empirical rate is about
$1.9\%$ on this corpus. Diagnosis reaches \MatrixSpectraDiag{}: $156$ of
$158$ different pairs recover both verdict and root cause
(Table~\ref{tab:rootcause}). The two misses are
conservative \textsc{ChangedSemanticBehavior} fallbacks on deleted
initializations masked by another write; deep-composition diagnosis is
$51/51$.
Both misses share one abstraction: the read-before-write summary is
order-insensitive, so deleting an initialization from one script is masked
when another script writes the same variable. They are generic fallbacks
with correct verdicts. The composed stratum causes no additional
diagnostic loss because canonicalization strips the transform stack before
classification.

\begin{table}[t]
\centering
\caption{Diagnosis by injected root cause: pairs whose verdict and root
cause are both recovered, over all validated different-labeled pairs.}
\label{tab:rootcause}
\small
\begin{tabular}{lrr}
\toprule
\textbf{Injected root cause} & \textbf{Pairs} & \textbf{Recovered}\\
\midrule
TriggerChange & 25 & 25\\
ExtraKillEdge & 26 & 26\\
ValueChange & 22 & 22\\
EffectRemoved & 18 & 18\\
GuardChange & 16 & 16\\
ExtraJoinEdge & 15 & 15\\
FramePathChange & 11 & 11\\
BroadcastEdgeRemoved & 7 & 7\\
ChangedFrameBoundary & 7 & 7\\
ChangedCloneMultiplicity & 5 & 5\\
UninitializedRead & 4 & 2\\
MissingJoinEdge & 2 & 2\\
\midrule
Total & 158 & 156\\
\bottomrule
\end{tabular}
\end{table}

\begin{table}[t]
\centering
\caption{Verdict quality on the mutation corpus under strict scoring.
Overall accuracy counts an abstention as an error; Acc@dec and Coverage
give the abstention-aware decomposition. FalseEq counts different-labeled
pairs classified as equivalent, the failure mode that certifies defective
programs. Diagnosis requires the verdict and the
injected root cause together; methods that emit no root causes show --.}
\label{tab:matrix}
\small
\setlength{\tabcolsep}{4.5pt}
\begin{tabular}{lrrrrr}
\toprule
\textbf{Method} & \textbf{Overall} & \textbf{Acc@dec} & \textbf{Cov.} & \textbf{FEq} & \textbf{Diag.}\\
\midrule
\tool{} & \MatrixSpectraOverall & \MatrixSpectraAccuracy & \MatrixSpectraCoverage & \MatrixSpectraFalseEq & \MatrixSpectraDiag\\
\quad static kernel & \MatrixSpectraOverall & \MatrixSpectraAccuracy & \MatrixSpectraCoverage & \MatrixSpectraFalseEq & \MatrixSpectraDiag\\
\quad without POR & \MatrixSpectraNoporOverall & \MatrixSpectraNoporAccuracy & 1.000 & \MatrixSpectraNoporFalseEq & \MatrixSpectraNoporDiag\\
\quad \lfinal{} only & \MatrixSpectraFinalOverall & \MatrixSpectraFinalAccuracy & 1.000 & \MatrixSpectraFinalFalseEq & --\\
Fingerprint equality & \MatrixBHashOverall & \MatrixBHashAccuracy & 1.000 & \MatrixBHashFalseEq & --\\
Opcode multiset & \MatrixBOpcodeOverall & \MatrixBOpcodeAccuracy & 1.000 & \MatrixBOpcodeFalseEq & --\\
Opcode trace & \MatrixBTraceOverall & \MatrixBTraceAccuracy & 1.000 & \MatrixBTraceFalseEq & --\\
Tree distance & \MatrixBStructOverall & \MatrixBStructAccuracy & 1.000 & \MatrixBStructFalseEq & --\\
Dynamic ($N{=}5$) & \MatrixBDynfiveOverall & \MatrixBDynfiveAccuracy & 1.000 & \MatrixBDynfiveFalseEq & --\\
GLM-5.1 & \MatrixLlmGlmOverall & \MatrixLlmGlmAccuracy & \MatrixLlmGlmCoverage & \MatrixLlmGlmFalseEq & \MatrixLlmGlmDiag\\
Qwen3.6-plus & \MatrixLlmQwenOverall & \MatrixLlmQwenAccuracy & \MatrixLlmQwenCoverage & \MatrixLlmQwenFalseEq & \MatrixLlmQwenDiag\\
Kimi K2.6 & \MatrixLlmKimiOverall & \MatrixLlmKimiAccuracy & \MatrixLlmKimiCoverage & \MatrixLlmKimiFalseEq & \MatrixLlmKimiDiag\\
\bottomrule
\end{tabular}
\end{table}

\subsection{RQ3: Baseline Comparison}

Every family fails in its predicted direction. Structural comparison avoids
false equivalence and rejects refactorings ($182$ false differences).
Opcode abstractions accept mutants with unchanged opcode statistics
(\MatrixBHashFalseEq{}--\MatrixBTraceFalseEq{} false equivalences). Dynamic
testing reaches \MatrixBDynfiveOverall{} overall, certifies
\MatrixBDynfiveFalseEq{} defective programs, and costs two orders of
magnitude more time. LLMs form a coverage--soundness frontier: Kimi~K2.6 is
strongest (\MatrixLlmKimiOverall{} overall,
\MatrixLlmKimiCoverage{} coverage) with
\MatrixLlmKimiFalseEq{} false equivalences; GLM-5.1 and Qwen3.6-plus also
commit false equivalences and trail \tool{} on diagnosis. McNemar tests
reject every baseline and ablation at $p<0.01$.
Abstention is directional: GLM-5.1 abstains on $17$ equivalent pairs against
$2$ different ones, and Qwen3.6-plus on $38$ against $13$. The models often
point at a difference; the missing capability is exhaustive equivalence
certification. Across all eleven comparisons, every pair on which \tool{}
and another method disagreed in correctness was decided correctly by
\tool{}.

Table~\ref{tab:strata} isolates deep composition, where one defect hides
among benign refactorings. Structural comparison collapses from
\MatrixBStructSingleOverall{} to \MatrixBStructHardOverall{}; dynamic testing
concentrates \MatrixBDynfiveHardFalseEq{} false equivalences there. Kimi~K2.6
and GLM-5.1 hold accuracy better than Qwen3.6-plus, and all three still
certify defective programs. \tool{} remains $1.000$ because canonicalization
removes the transform stack before classification.

\begin{table}[t]
\centering
\caption{Overall accuracy by stratum (abstention counts as error), and
false equivalences on the deep-composition stratum alone.}
\label{tab:strata}
\begin{tabular}{lrrr}
\toprule
\textbf{Method} & \textbf{Single-edit} & \textbf{Composed} & \textbf{FEq (comp.)}\\
\midrule
\tool{} & \MatrixSpectraSingleOverall & \MatrixSpectraHardOverall & \MatrixSpectraHardFalseEq\\
Tree distance & \MatrixBStructSingleOverall & \MatrixBStructHardOverall & --\\
Dynamic ($N{=}5$) & -- & \MatrixBDynfiveHardOverall & \MatrixBDynfiveHardFalseEq\\
GLM-5.1 & \MatrixLlmGlmSingleOverall & \MatrixLlmGlmHardOverall & \MatrixLlmGlmHardFalseEq\\
Qwen3.6-plus & \MatrixLlmQwenSingleOverall & \MatrixLlmQwenHardOverall & \MatrixLlmQwenHardFalseEq\\
Kimi K2.6 & \MatrixLlmKimiSingleOverall & \MatrixLlmKimiHardOverall & \MatrixLlmKimiHardFalseEq\\
\bottomrule
\end{tabular}
\end{table}

\subsection{RQ4: Targeted Scenarios on the Ambiguous Bucket}

The ambiguous bucket contains \MatrixRqfourBucket{} mutants left unresolved
by random scenarios, concentrated in untriggered effects, weak join
contention, frame timing, and receiver fallout. The static kernel records
typed conditional frontiers for \MatrixRqfourStaticDiff{} of them; targeted
scenarios confirm \RqFourExposed{} concrete divergences (rate
\RqFourRate{}), usually on the first synthesized scenario, including
position, costume, visibility, and clone-count witnesses. The dynamic
baseline certifies eleven of these fourteen as equivalent. The unexposed
remainder mixes genuinely equivalent mutants and divergences below the
current oracle surface, so conditional frontiers remain conditional.
The exposed witnesses use fresh seeds disjoint from both the labeling battery
and the dynamic baseline. Three of the fourteen come from trigger-directed
scenarios that press the keys
named in the typed diff. The remaining $74$ bucket members include trigger
changes with no observable effect, join edits whose receivers finish too
quickly to contend, broken broadcasts without sampled fallout, and value or
guard edits below the current oracle signal.

\subsection{RQ5: Ablations and the Lens Matrix}

Both ablations fail as predicted. Removing POR
(\MatrixSpectraNoporOverall{} overall), all twenty errors are false
differences on commuted independent writes. The \lfinal{}-only configuration
(\MatrixSpectraFinalOverall{}) keeps 1.000 precision and drops recall to
$0.215$ because $124$ event-, frame-, and effect-level divergences are
invisible to final state. On all $243$ pairs with explicit \lfinal{} labels,
\lfinal{} verdicts match the label vector, showing that each verdict is tied
to its observer.
A glide-for-jump pair illustrates the point: its label vector is
\[
\langle \ldefault{}{:}\,\textsc{Diff},\lfinal{}{:}\,\textsc{Eq}\rangle .
\]
The full kernel reports \textsc{FramePathChange} under \ldefault{}, while the
\lfinal{} configuration proves equivalence through final-state transfer. A
single-bit checker has no observer field to express both verdicts.

\subsection{RQ6: Cost and Escalation Profile}

The static kernel decides the validated corpus without VM execution at a
$0.22$s median. All $286$ equivalences close by direct canonical equality;
no headline equivalence relies on finite execution, \lfinal{} transfer, SMT
discharge, or bounded tie-group acceptance. Different-labeled pairs close by
static root-cause evidence; residual obligations appear in the ambiguous
bucket, while the controlled suite exercises the \lfinal{} transfer path
(Table~\ref{tab:paths}). The tail ($5.0$s p90,
$27$s p99) comes from
minimal-delta explanations on composed differences. The dynamic baseline
spends $29$s median and still has \MatrixBDynfiveFalseEq{} false
equivalences; \tool{} reserves VM cost for label validation and RQ4
frontiers. LLM latency is provider-bound at $4$--$10$s.

\begin{table}[t]
\centering
\caption{Decision-path profile. Headline counts are conclusive corpus
verdicts; frontier counts are conditional records in the ambiguous bucket;
controlled counts are conclusive outcomes in the 26-case semantic suite.}
\label{tab:paths}
\scriptsize
\setlength{\tabcolsep}{3pt}
\begin{tabular}{lrrr}
\toprule
\textbf{Path} & \textbf{Headline} & \textbf{Frontier} & \textbf{Controlled}\\
\midrule
Canonical equality & 286 & 0 & 7\\
Tie-group bijection & 0 & 0 & 0\\
\lfinal{} transfer & 0 & 0 & 1\\
Static root-cause fact & 158 & 88 cond. & 18\\
Targeted VM witness & 0 & 14 & 0\\
\textsc{Unknown}/frontier & 0 & 74 & 0\\
\bottomrule
\end{tabular}
\end{table}

\subsection{Controlled Phenomenon Coverage}

Independent of the mutation corpus, a hand-constructed suite of
\SpectraSemanticCases{} minimal pairs covers each taxonomy phenomenon and the
main static acceptance paths: 7 canonical equivalences, 1 \lfinal{} transfer,
and 18 static-difference verdicts. \tool{} decides all of them correctly with
root-cause accuracy \SpectraRootCauseAccuracy{}. VM-witness cases appear in the
frontier bucket; the controlled suite counts conclusive static outcomes.
Renderer and real-project smoke runs
serve as implementation validations outside the effectiveness claims.

\section{Discussion}
\label{sec:discussion}

\emph{Oracle transport.} Execution-based equivalence needs a transported
oracle: stimuli translate forward through an identifier bijection and
observations translate back. Two development failures, broadcast case
normalization and renamed-message stimuli, came from violating this rule.

\emph{Ambiguity and deployment.} The \MatrixRqfourBucket{} ambiguous members
mix genuine equivalents, unobserved divergences, and longer-horizon
behaviors. RQ4 keeps them as typed frontiers until targeted scenarios expose
\RqFourExposed{} concrete divergences. In tutoring, \textsc{Unknown} routes to
review, while false equivalence ends debugging incorrectly. Strict scoring and
explicit unknowns match the deployment contract. The static kernel handles
interactive cases at a $0.22$s median; unknowns route to dynamic evidence or
human review.

\emph{Artifact boundary.} The public review path exposes the semantic checker
through a fixed interface: CSIR export, canonical comparison, controlled-suite
drivers, manifest-based \tool{} reruns, trace replay, and aggregation. The
private tutor service supplies deployment glue, credentials, and UI
integration. Those components do not affect the verdict relation or the
reported corpus metrics. This boundary lets reviewers rerun the analysis
layer without gaining access to production infrastructure.

\section{Threats to Validity}
\label{sec:threats}

\emph{Construct and internal validity.} We treat the reference VM as the
operational semantics for Scratch 3 behavior in this study, but
instrumentation can still be wrong. The harness records
concrete execution only, never calls the static checker for labels, and uses
disjoint seed namespaces. Generator/classifier coupling remains possible
because operators share the taxonomy; prior bug patterns, deep-composition
pairs, VM witness admission, and quarantine reduce this risk.

\emph{Selection, external, and LLM validity.} Witness admission
under-represents rare-schedule and long-horizon bugs; 0/158 FEq is a
corpus-specific empirical result. The corpus is Scratch 3 from
one crawl source; other block languages need new carriers, cuts, and lenses.
Public seeds may appear in LLM training data and served models drift, so the
artifact pins model ids, prompts, raw responses, and parsed JSON.
The $600$-block cap removes projects whose traces are dominated by assets,
long loops, or renderer effects; it keeps corpus generation reproducible and
interactive, while under-sampling very large games and animation-heavy
projects. The excluded $26$ seeds are recorded in the manifest audit trail, so
future corpora can raise the cap and measure whether clone-, timer-, and
renderer-heavy programs change the frontier mix.

\section{Related Work}
\label{sec:related}

\emph{Scratch analysis and testing.} Hairball~\cite{boe2013hairball},
Dr.~Scratch~\cite{morenoleon2015drscratch}, LitterBox~\cite{fraser2021litterbox,
fraedrich2020bugs}, Bastet~\cite{stahlbauer2020verified}, Whisker-style
testing~\cite{stahlbauer2019testing,deiner2023testgen,goetz2022model},
NuzzleBug~\cite{deiner2024nuzzlebug}, and test-based hinting
~\cite{obermueller2021catnip} analyze or guide one program. \tool{} compares
two programs and ties each root cause to a carrier and observation lens.
Recent Scratch feedback systems~\cite{li2026raven,si2025stitch},
repair work~\cite{si2026ecoscratch}, and multimodal
benchmarks~\cite{si2025viscratch,si2026scratcheval} use LLMs, gameplay videos,
and execution evidence to surface semantic bugs that block structure alone can
miss. \tool{} targets the complementary equivalence problem needed to compare
learner programs, reference solutions, and candidate repairs.

\emph{Differencing and concurrency.} Prior differencing and
equivalence tools~\cite{falleri2014gumtree,fluri2007changedistiller,
zhang1989ted,jackson1994semanticdiff,lahiri2012symdiff,person2008differential,
godlin2009regression,ramos2011equivalence,pnueli1998translation,roy2009clones}
target syntax or sequential programs; Scratch adds hats, cooperative
scheduling, broadcasts, ask queues, clones, and renderer-visible state. The
trace quotient builds on Mazurkiewicz theory~\cite{mazurkiewicz1987trace} and
partial-order reduction~\cite{godefroid1996partial,flanagan2005dpor,
valmari1998state}; CEGAR~\cite{clarke2003cegar} handles frontiers, and
lenses make the observer explicit~\cite{milner1989communication}.
Work on semantic merge conflicts and feedback generation for conventional
code also compares related program versions or candidate repairs, but targets
textual languages and judge-style traces rather than event-driven Scratch
lenses~\cite{zhang2022mergeconflicts,zhang2022competitionfeedback}.

\emph{Mutation methodology and artifacts.} The evaluation adapts mutation
analysis~\cite{demillo1978mutation,jia2011mutation} to a setting where
equivalent mutants are expected, useful, and lens-dependent. Traditional
mutation studies often discard equivalent mutants as noise~\cite{schuler2013equivalent};
we keep unresolved cases as an ambiguous bucket, attach typed frontiers, and
separate random validation from targeted exposure. The released manifests
preserve the operator chain, seed identity, checksums, VM scenarios, verdicts,
and aggregation inputs, making the benchmark auditable even when a reviewer
chooses to rerun only a controlled suite or sample.

\section{Conclusion}
\label{sec:conclusion}

Behavioral equivalence for event-driven block programs is lens-parametric.
\tool{} implements that contract with CSIR, trace normal forms over typed
footprints, rename-invariant canonicalization, content-addressed procedures,
typed witnesses, unknown frontiers, and VM-backed refinement. On
\MatrixUsablePairs{} validated pairs, including $158$ VM-witnessed
differences, it makes $0/158$ false-equivalence claims and recovers root
causes with \MatrixSpectraDiag{} accuracy. The broader lesson is
methodological: learner-program equivalence
needs an explicit observer, a transported test oracle, strict scoring, and a
place for unknowns. Future work widens the frontier with wall-clock
virtualization, monitor/audio oracles, procedure summaries, renderer
witnesses, and front ends for other block languages.

\clearpage
\bibliographystyle{IEEEtran}
\bibliography{references}

\end{document}